\documentclass[twocolumn,10pt, aps]{revtex4-2}%
\usepackage[paperwidth=210mm,paperheight=297mm,centering,hmargin=2cm,vmargin=2.5cm]{geometry}
\usepackage{amsfonts}
\usepackage{amsmath}
\usepackage{amssymb}
\usepackage{bm}
\usepackage{graphicx}
\usepackage{color}
\usepackage{soul}
\usepackage{epsfig}%
\usepackage[dvipsnames]{xcolor} 
  \definecolor{bleu_cite}{RGB}{0,0,255}
\usepackage[colorlinks=true,allcolors = black,citecolor=bleu_cite,  ]{hyperref} 
\usepackage{siunitx}

\begin{document}

\title{\textcolor{black}{Superlattice quantum solid of dipolar excitons}} 

\author{Camille Lagoin$^{1,2}$, Kirk  Baldwin$^3$, Loren Pfeiffer$^3$ and Fran\c{c}ois Dubin$^{1,2}$}
\affiliation{$^1$ CRHEA, CNRS and Université Côte d'Azur, Valbonne, France }
\affiliation{$^2$ Institut des Nanosciences de Paris, CNRS and Sorbonne Universit{\'e}, Paris, France}
\affiliation{$^3$ PRISM, Princeton Institute for the Science and Technology of Materials, Princeton University, Princeton, USA}

\begin{abstract}
\textcolor{black}{We study dipolar excitons confined at 330 mK in a square electrostatic lattice of a GaAs double quantum well. In the dipolar occupation blockade regime, at 3/2 filling, we evidence  that excitons form a  face-centred superlattice quantum solid. This phase is realised with high purity across 36 lattice sites, in a regime where the excitons mean interaction energy exceeds the depth of the lattice confinement. The superlattice solid then closely relates to Wigner crystals.}
\end{abstract}

\maketitle

\textcolor{black}{\textbf{Introduction} Strongly correlated quantum particles confined in lattice potentials are specifically described by the Hubbard Hamiltonian. This model predicts that nearest neighbour (NN) couplings greatly enrich the texture of many-body ground states. Notably, extended interactions favour density waves spontaneously breaking translational lattice symmetry at specific fractional lattice fillings, for instance stripes or checkerboard (CB) patterns. Such quantum solids have been extensively studied for fermionic systems \cite{wise2008charge,fradkin2015colloquium,regan2020mott,xu2020correlated,huang2021correlated,jin2021stripe}. They have also been theoretically predicted for dipolar bosons \cite{Goral2002,capogrosso2010quantum,Baranov2012,Dutta2015}, and accessed recently with ultra-cold atoms \cite{Greiner2023} and dipolar excitons \cite{lagoin2022extended} of GaAs double quantum wells. }

\textcolor{black}{Dipolar excitons confined in gate-defined electrostatic lattices provide a unique platform to study Bose-Hubbard (BH) physics in the solid-state \cite{Zeng2023,Xiong2023,Park2023}. For lattice periods greater than around 400 nm, excitons explore the regime restricted to on-site dipolar repulsions.  Mott insulators (MIs) with exactly one or two excitons in every site have then been observed \cite{lagoin2021mott}. For lattice periods less than around 250 nm, one enters the regime controlled by NN interactions yielding an extended BH description. At the same time, on-site dipolar repulsions exceed the lattice depth and the system evolves in a dipolar blockade regime where lattice sites have occupations bound to unity. In this situation, MI and CB solids have been reported at unity and half-filling respectively \cite{lagoin2022extended}. Yet, for dipolar bosons the structure of the phase diagram remains unexplored for average lattice fillings $\bar{\nu}$ beyond unity.}

\textcolor{black}{In this Letter, we show that, in the occupation blockade regime, dipolar excitons realise a superlattice quantum solid at $\bar{\nu}=3/2$. This superlattice is obtained by preparing a MI in a gate-defined electrostatic potential, with one exciton per site ($\bar{\nu}=1$). For larger occupations the MI induces a dipolar lattice confining additional carriers. At $\bar{\nu}=3/2$, a quantum insulator emerges, made by a MI and a CB solid in the electrostatic and dipolar potential respectively. Strikingly the excitons mean repulsion energy then exceeds the depth of the electrostatic confinement, marking that the superlattice solid is reminiscent of Wigner crystals.}

\textbf{Dipolar lattice}  As reported in recent works\,\cite{lagoin2022extended,lagoin2023dual}, we imprint a 250 nm period square electrostatic lattice in the plane of a GaAs double quantum well. \textcolor{black}{The latter confines electrons and holes realising dipolar excitons trapped in the regions where the applied field is the largest. Experimentally, dipolar excitons are optically injected in a region extending across around (8x8) $\mathrm{\mu m}^2$ (1000 sites), with a stretched gaussian laser excitation which intensity controls $\bar{\nu}$ (see Supplementals)}. In the electrostatic lattice, repulsive NN dipolar interactions are dominant at our experimental temperature \textcolor{black}{$T=330$ mK}. Hence, we implement the extended BH Hamiltonian diagnosed by probing the spatially and spectrally resolved photoluminescence (PL) that excitons emit at thermal equilibrium, long after optical injection. Figure 1.a illustrates that the lattice depth (around 250 $\mu$eV) is small compared to the strength of on-site dipolar repulsions $U$ ($\sim1$ meV). \textcolor{black}{Hence, our device evolves deep in the dipolar occupation blockade regime.}

\textcolor{black}{Figure 1.c displays the PL spectrum radiated by the electrostatic lattice at $\bar{\nu}=1$. The profile is quantitatively adjusted by a single emission line -- labelled $(0)$ -- with a width given by our instrumental resolution (red area in Fig.1.c).} This manifests that excitons all occupy the same energy level (Fig.1.a). \textcolor{black}{Following the fluctuation dissipation theorem \cite{Gemelke2009,{lagoin2021mott}}, we monitored statistical fluctuations of the PL intensity to quantify the exciton compressibility $\kappa= \frac{\partial \bar {n}}{\partial \mu}$, $\mu$ denoting the chemical potential and $\bar{n}$ the mean occupation of the occupied energy state (Supplementals). For a classical phase, $\kappa$ is given by the level of poissonian fluctuations while sub-poissonian values signal a quantum insulator. Hence, we verify that the experiments shown in Fig.1.c reveal a MI since we measure that the compressibility normalised to poissonian fluctuations is reduced to $\kappa=0.54$. Note that here and in the following $\kappa$ refers to normalised compressibility.}

\textcolor{black}{Figure 1.d shows the PL spectrum obtained for a laser excitation yielding $\bar{\nu}=1.2$ (Supplementals). We note that the spectrum broadens towards higher energies compared to Fig.1.c. This manifests that Mott order is suppressed in the electrostatic potential \cite{lagoin2021mott,lagoin2022extended}, as confirmed by the measured classical  compressibility $\kappa=1.1$.} Moreover, new emission lines emerge at around 600 $\mu$eV higher energy (blue in Fig.1.d). To interpret these contributions, we first note that the dipolar repulsion strength between excess excitons and the ones confined in the electrostatic lattice varies periodically (see purple area in Fig.1.a). It decreases from around $U$ when excess excitons have a wave-function centred at the position of singly occupied electrostatic lattice sites, to about 600 $\mu$eV at the location of electrostatic barriers (Supplementals). Excess excitons are thus confined in a dipolar potential with around 500 $\mu$eV depth (red wave in Fig.1.b). One theoretically expects two confined energy levels in this potential, separated by approximately 200 $\mu$eV. \textcolor{black}{These are referred to as Wannier states -- WS1 and WS2 and labelled as (I) and (III) in Fig.1.c-d}. The PL spectrum shown in Fig.1.d confirms the structure of the dipolar confinement. \textcolor{black}{It is characterized by contributions from both WS so that the profile is adjusted by only setting their occupations (blue curves and inset).}

\textcolor{black}{\textbf{Inter-lattice interactions} Given the geometrical arrangement of the two lattices \textcolor{black}{(electrostatic and dipolar)}, inter-lattice NN dipolar repulsions have the largest magnitude $V_{E,D}$}. Indeed, corresponding sites have a separation reduced to 177 nm. Theoretically, \textcolor{black}{$V_{E,D}$} then lies around 90 $\mu$eV whereas intra-lattice NN interactions have amplitudes, \textcolor{black}{$V_{E,E} \equiv V_{E}$ and  $V_{D,D} \equiv V_{D}$} for the electrostatic and dipolar potentials respectively, about 30 $\mu$eV (see Ref.\cite{lagoin2022extended} and Supplementals). \textcolor{black}{Figure 1.b illustrates that excitons confined in the electrostatic lattice have then five accessible energy states, well separated by $V_{E,D}$ and labelled (0) to (4) to reflect the corresponding number of inter-lattice NN interactions. On the other hand, in the dipolar lattice accessible energies have a separation reduced to $V_{D}$. Remarkably intra- and inter-lattice interactions structure the PL spectrum that then reveals spatial order in our device, as shown below.}

\textcolor{black}{The magnitude ($V_{E,D}$) of inter-lattice repulsions is well confirmed by the spectral profile displayed in Fig.1.d. Indeed, its low energy part (red) maps accessible energy states in the electrostatic lattice. It is quantitatively reproduced by adjusting the contributions from (0) to (3) NN couplings (inset). The fraction of single interactions (line (1)) is twice larger than the one for 2NN (line (2)),  indicating as expected that adjacent sites of the dipolar potential are weakly occupied together at low fillings. Let us then note that by modelling the PL spectral profile as in Fig.1.d we extract the average distribution of occupied sites in the dipolar potential. Thus we partially overcome our optical instrumental resolution of (1.5x1.5) $\mu$m$^2$ corresponding to 36 lattice sites.}

\textbf{Dual Mott and checkerboard orders} 
\textcolor{black}{The middle panel of Fig.2.a shows the spatially and spectrally resolved PL when the average filling of our device is set to $\bar{\nu}\sim3/2$. Figure 2.b reveals that the spectral profile is governed by two emission lines at the center of the illuminated region (zone C). Its low energy component marks the contribution from excitons confined in the electrostatic potential. It lies at the energy for exactly 2 inter-lattice NN interactions (line (2)). On the other hand, the higher energy emission is due to excitons confined in the dipolar lattice, at the energy of the first WS with no intra-lattice NN interactions (line (I)). These combined emissions signal that excitons realise a face centred square superlattice. The top panel of Fig.2.a illustrates that this phase consists of a MI in the electrostatic potential and a CB in the dipolar one. Figure 2.b actually reveals that the measured spectral profile exhibits a 75$\%$ overlap with the spectrum theoretically expected for such superlattice quantum solid.} The degree of purity is mostly limited by the fraction of excitons occupying the second WS in the dipolar lattice \textcolor{black}{($\sim20\%$ -- inset Fig.2.b)}. Let us then underline that our experiments rely on stroboscopic accumulations averaging millions of realisations (Supplementals). The inset of Fig.2.b then shows that in the electrostatic lattice an average of 80$\%$ of excitons experience 2NN interactions with the \textcolor{black}{dipolar lattice}, while the mean fractions of excitons with 0 and 3 NN interactions are both reduced to about 10$\%$. These amplitudes highlight that irregularities of the CB pattern are in average bound to a single excess exciton/vacancy across 36 sites.

We further studied the exciton compressibility in both electrostatic and dipolar lattices, \textcolor{black}{$\kappa_E$} (Fig.2.d) and \textcolor{black}{$\kappa_{D}$} (Fig.2.e) respectively. These are obtained by monitoring fluctuations of the corresponding maxima of the PL spectrum (Supplementals). Figures 2.d-e show that \textcolor{black}{$\kappa_{E}$}  and \textcolor{black}{$\kappa_{D}$} are both minimised with sub-poissonian values at the center of the laser-excited region (gray intensity profile). These variations signal that quantum insulators are realised, across around 50 sites probed along the vertical direction of the lattice (darker red and blue bars). Thus we verify directly that Fig.2.b reveals combined Mott and CB quantum solids, \textcolor{black}{while we observe that on the sides of the optical injection excitons return to compressible phases.}

\textbf{Interlocked quantum insulators} \textcolor{black}{Figure 2.c shows the PL spectrum emitted slightly off the center of the optically excited zone (S in Fig.2.a). Compared to Fig.2.b, the PL from the electrostatic lattice varies weakly (see insets) reflecting that 2NN inter-lattice interactions are dominant. By contrast, the PL emitted by the dipolar lattice is strikingly dominated by a contribution (line II) at around 80 $\mu$eV higher energy than WS1 (I). This signals that a large fraction of excitons experience around 2 intra-lattice NN interactions. Hence, the average spatial distribution of occupied dipolar lattice sites mixes stripes and CB like domains in this adjacent 36 sites cluster. Figure 2.a (bottom panel) illustrates such situation for which we measure a classical compressibility (see Fig.2.d-e). Hence, differences between the spectra shown in Figs. 2.b and 2.c reveal that non-periodic fillings of the dipolar lattice suppress global quantum order.}

\textcolor{black}{In our device we expect that quantum insulating states are only accessed when excitons confined in the electrostatic lattice all interact with the same distribution of occupied dipolar lattice sites}. Otherwise, Fig.3.a illustrates that inter-lattice NN interactions itinerantly vary the energy of the electrostatic lattice sites, such that the MI is suppressed in a fashion reminiscent of the Bose-glass transition \cite{Fallani2007,Maciej2003,deMarco2016}. For our square potential this requirement is fulfilled for two quantum solid phases, at $\textcolor{black}{\bar{\nu}}=5/4$ and 3/2, for which excitons in the electrostatic lattice all experience 1 or 2 inter-lattice NN interactions respectively (Fig.2.a illustrates the superlattice conformation \textcolor{black}{at $\bar{\nu}=3/2$}).

We probed \textcolor{black}{$\kappa_E$} just above total unitary filling to verify that disordered distributions of excitons in the \textcolor{black}{dipolar lattice suppress quantum ordering}. \textcolor{black}{Precisely, we varied \textcolor{black}{$\bar{\nu}$} with the  laser excitation power} and modelled PL spectra to extract the fraction of excitons that experience at least 1 inter-lattice NN interaction \textcolor{black}{(Supplementals)}. \textcolor{black}{Figure 3.b highlights that $\kappa_E$ increases steeply when over around 25$\%$ of excitons in the electrostatic potential experience at least one inter-lattice NN interaction}. For these experiments averaging around 50 sites, this threshold corresponds to a mean number of only 3 excitons in the dipolar lattice. \textcolor{black}{Beyond, the exciton compressibility returns to classical magnitudes.}

To further probe inter-lattice correlations we measured \textcolor{black}{$\kappa_{E}$ and $\kappa_{D}$ for a wide range of $\bar{\nu}$}. Figure 3.d reveals that combined quantum order is only realised at $\bar{\nu}$ around 3/2.  \textcolor{black}{Indeed at $\bar{\nu}=5/4$ intra-lattice second NN interactions are too weak to favour a quantum insulating phase at $T=330$ mK} \cite{capogrosso2010quantum}. \textcolor{black}{Moreover, Fig.3.d shows that for $\bar{\nu}\neq3/2$ the exciton compressibility lies well above the level of poissonian fluctuations}. For instance \textcolor{black}{$\kappa_E$ and $\kappa_{D}$} are about 1.5 at $\bar{\nu}\sim$ 5/4. \textcolor{black}{These amplitudes contrast with the ones for $\bar{\nu}\sim$ 1 (Fig. 3.b) where $\kappa_E$ is around unity in the classical regime}. We attribute this difference as the result of a strong \textcolor{black}{positive feedback} between occupation fluctuations in the two lattices, \textcolor{black}{which then fragilises the superlattice insulator}. This channel is theoretically expected \cite{ortner2009quantum,pupillo2008cold} and certainly also limits the minimum compressibility at $\bar{\nu}=3/2$.

\textcolor{black}{Finally, we note that for $\bar{\nu}\sim9/8$ excitons in the electrostatic lattice sites experience in average 1 NN inter-lattice repulsion (see Fig.S2). Their mean energy is then increased by $V_{E,D}$, and becomes comparable to the lattice depth, since 4 intra-lattice NN interactions also contribute (with magnitude $4V_E$ -- Fig.1.a).}  At \textcolor{black}{$\bar{\nu}$ above 4/3}, the mean energy is further increased by 1 inter-lattice NN interaction so that excitons reach energy levels that are theoretically no longer confined by the electrostatic potential. Figure 3.c illustrates this situation where the PL spectrum nevertheless consists of two individual lines (see Fig.2.b). This profile evidences that excitons are spatially ordered with sub-possonian compressibility. \textcolor{black}{We then deduce that the occupation of the dipolar potential stabilises the underlying MI phase (green in Fig.3.c), no longer trapped by the sole electrostatic potential. The superlattice solid at $\bar{\nu}\sim3/2$ then closely ressembles self-assembled quantum crystals \cite{pupillo2008cold,ortner2009quantum,Boronat2017}.}

\textbf{Conclusions} We have shown that strong dipolar repulsions allow assembling interlocked Mott and checkerboard phases \textcolor{black}{to form a superlattice insulator}. Although, we rely on an shallow electrostatic lattice to prepare a MI, global quantum order emerges in a parameter space where the lattice confinement depth is lower than the strength of dipolar repulsions. This regime is reminiscent of Wigner crystallisation which is actually theoretically accessible for dipolar excitons of GaAs double quantum wells \cite{Lozovik1998,DasSarma2006,filinov2010berezinskii}. Extending our studies to such spontaneously assembled quantum crystals would pave the way towards new frontiers of condensed-matter systems \cite{ortner2009quantum,pupillo2008cold}.

\section*{Acknowledgments}
We would like to thank D. Hrabovsky, S. Suffit and L. Thevenard for support, together with G. Pupillo for stimulating discussions, and M. Lewenstein, T. Grass, U. Bhattacharya and A. Imamo\u{g}lu for a critical reading of the manuscript. Our research has been financially supported by the French Agency for Research (contracts IXTASE and SIX). The work at Princeton University was funded by the Gordon and Betty Moore Foundation through the EPiQS initiative Grant GBMF4420, and by the National Science Foundation MRSEC Grant DMR 1420541.

\section*{Data availability}

Source data supporting all the conclusions raised in this manuscript are available for download upon request.

\section*{Financial interest}

The authors declare no competing financial interest.


\newpage

\onecolumngrid

\centerline{\includegraphics[width=\linewidth]{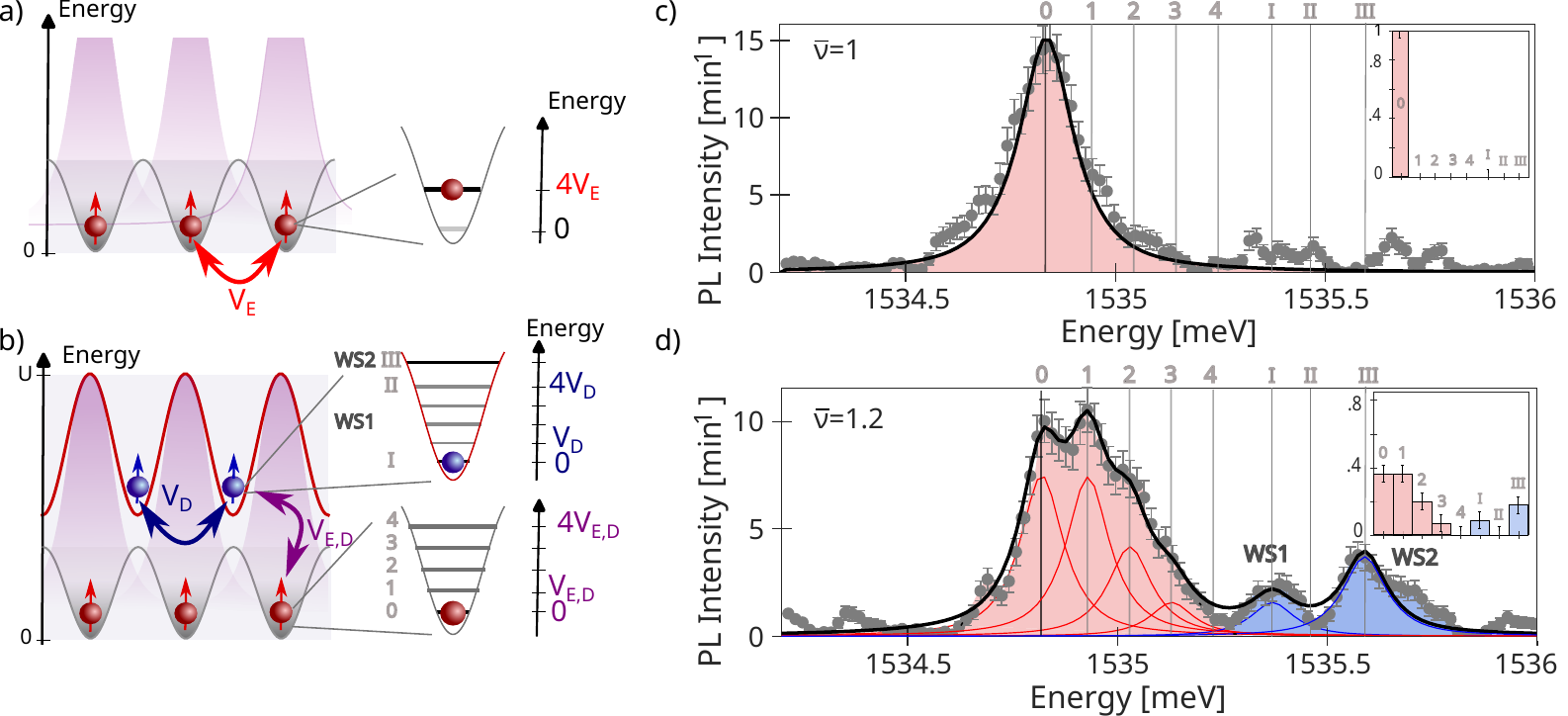}}
 \vspace{.5cm}\textbf{Fig. 1: }
\textcolor{black}{a) A MI at $\bar{\nu}=1$ (red balls in gray waves) imprints a dipolar potential (violet area) larger than the confinement depth. (Right) Excitons all lie at the same $4V_E$ energy due to 4 intra-lattice NN repulsions. b) Excitons (blue balls) confined in the dipolar lattice (red wave) induced by the MI. (Right) Exciton energies in the electrostatic potential are given by the number of inter-lattice interactions with strength $V_{E,D}$ (bottom). Intra-lattice interactions ($V_{D}$) control accessible energies in the dipolar lattice (top). c) PL spectrum radiated by a MI at unity filling, together with the profile of our spectral respone function (red area). The profile is adjusted (solid black line) by the sole line (0). d) PL spectrum at $\bar{\nu}=1.2$. The profile is adjusted (solid black line) by setting  the fractions of inter-lattice NN interactions (red areas for each number of NN -- (0) to (4) at energies given by vertical lines) and  the occupation of WS1 (I)  and WS2 (III) -- blue. In b) and c) insets show the occupation fraction of each accessible state. PL spectra are obtained by averaging across 36 lattice sites, error bars displaying our statistical precision. }
 \newpage
 
\centerline{\includegraphics[width=\linewidth]{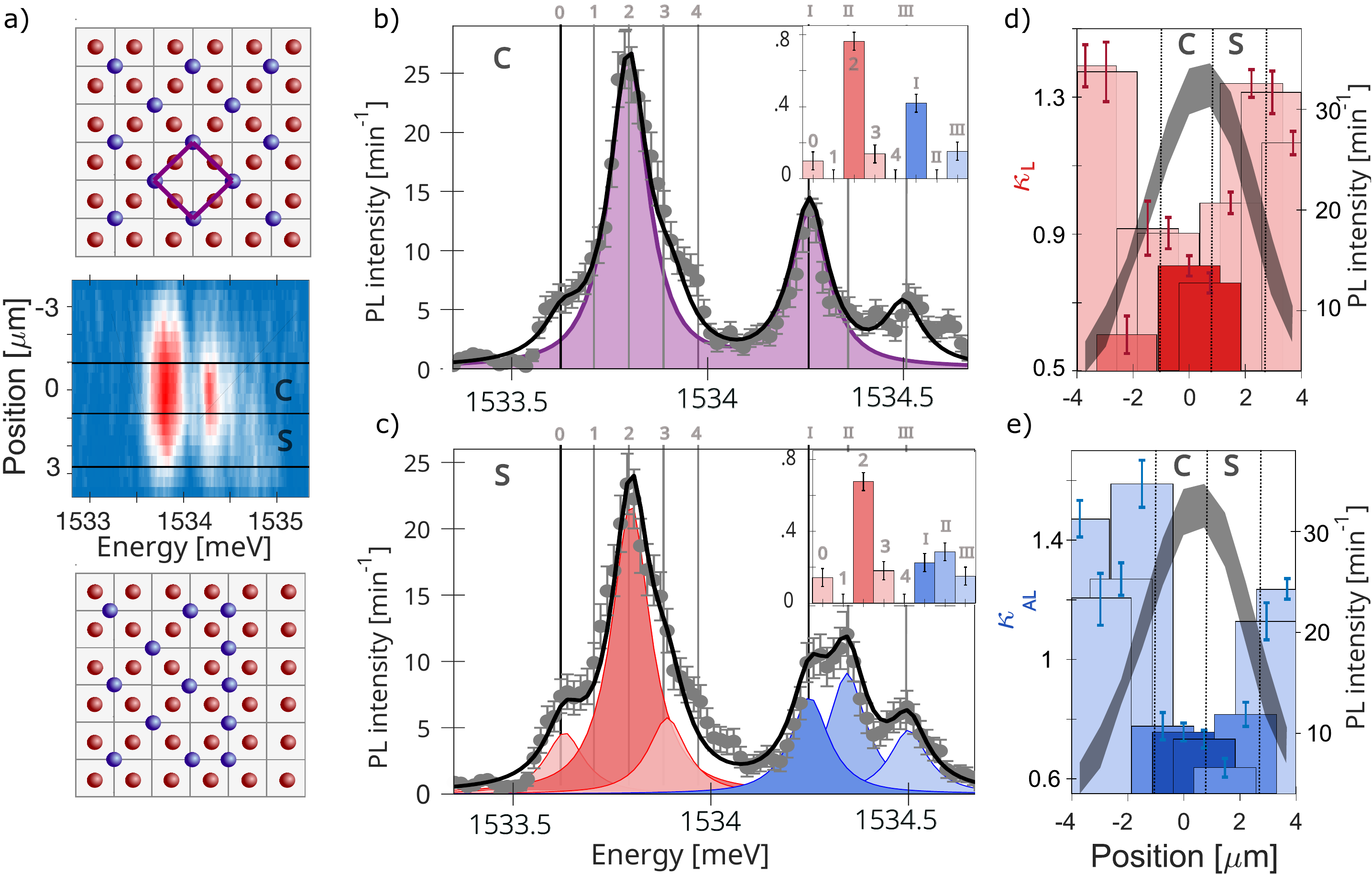}}
 \vspace{.5cm}\textbf{Fig. 2: }
\textcolor{black}{a) (top panel) Sketch of the superlattice quantum solid made by a CB in the dipolar lattice (blue) and a MI in the electrostatic one (red) across 36 sites. The violet square represents the elementary superlattice cell. (middle panel) Spatially and spectrally resolved PL at $\bar{\nu}=3/2$. (bottom panel) Disordered arrangement in the dipolar lattice around half-filling with stripes and CB patterns. b) PL spectrum emitted at the center of the illuminated region (C in panel a), together with the theoretical spectrum for a superlattice quantum solid (purple shaded area). c) PL spectrum measured in region S of panel a). In b)-c) PL spectra are averaged over 36 sites and model profiles (black lines) are obtained by adjusting the contributions from inter-lattice NN interactions (red bars in inset), and WS occupations together with NN interactions in the dipolar lattice (blue bars in inset). The insets in c) and d) show the occupation fractions of accessible states. d)-e) Vertically resolved exciton compressibility, averaged across 36 sites, in the electrostatic (d) and dipolar (e) lattices, together with the profile of the PL intensity (gray shaded area). In b)-e) error bars display our statistical precision. }
\newpage

\centerline{\includegraphics[width=.8\linewidth]{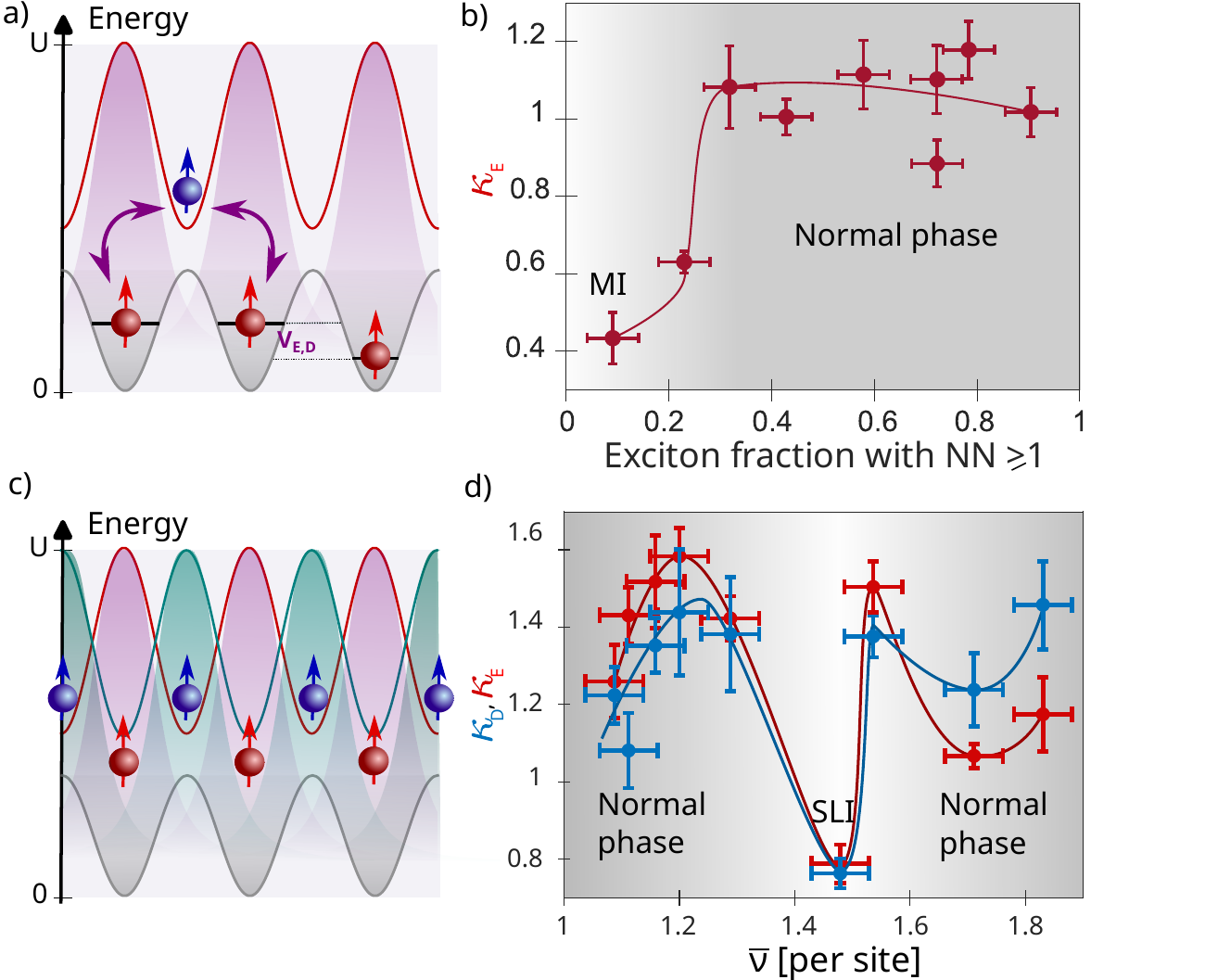}}
 \vspace{.5cm}\textbf{Fig. 3: }
\textcolor{black}{a) A single exciton in a dipolar lattice site (blue) shifts the energies of NN excitons in the electrostatic potential (red) by $V_{E,D}$. b) Exciton compressibility in the electrostatic lattice as a function of the fraction of sites with at least 1 NN in the dipolar lattice. When this observable reaches 0.85 we estimate that $\bar{\nu}\sim1.2$. c) The interaction potential created by excess excitons in the dipolar lattice (green area) dresses the electrostatic lattice confinement. Its depth is surpassed at $\bar{\nu}\gtrsim4/3$. d) Compressibilities $\kappa_E$ (red) and $\kappa_{D}$ (blue) as a function of the  total filling fraction $\bar{\nu}$ where SLI refers to superlattice insulator. In b) and d) measurements are all averaged across 50 lattice sites, error bars displaying our statistical precision while lines are guides for the eye.}

\onecolumngrid

\newpage

\centerline{\Huge{Supplementary Informations}}

\section*{Optical injection of dipolar excitons in a 250 nm period square lattice}

We study a field-effect device embedding two coupled 8 nm wide GaAs quantum wells separated by a 4 nm Al$_{.3}$Ga$_{.7}$As barrier \cite{lagoin2023dual,lagoin2022extended,lagoin2021mott}. This heterostructure was systematically probed at a temperature $T=$ 330 mK. Electronic carriers were optically injected by a laser excitation tuned at resonance with the direct exciton absorption of the two quantum wells. Electrons and holes then relax towards their respective minimum energy states, located in a distinct well since our heterostructure is biased by a DC voltage applied between the sample electrical ground and surface top electrodes. These gates are separated by 300 nm, quantum wells being positioned 200 nm under top electrodes, and a bias around ($-1$) V was applied. In this situation, Coulomb attraction between spatially separated electrons and holes leads to dipolar excitons characterised by a long lifetime ($\simeq$ 700 ns) and a record electric dipole (570 Debye), oriented perpendicular to the double quantum well plane.    

To emulate the Bose-Hubbard Hamiltonian, we patterned top gate-electrodes to imprint an electric-field spatially periodic in the plane of the GaAs double quantum well. The 250 nm period electrode square array was designed using finite element simulations which led to a pattern made by rectangular electrodes connected through thin wires (see Ref.\,\cite{lagoin2023dual,lagoin2022extended,lagoin2021mott} for more details). Dipolar excitons are high-field seekers and are then confined in the regions of the quantum wells placed under rectangular electrodes, since the applied electric-field is the strongest in this zone. Combined simulations and experimental probes allowed verifying that the electrostatic lattice depth is around 250 $\mu$eV\,\cite{lagoin2022extended}.

\textcolor{black}{
\section*{Modelling spectral profiles and density calibration}
With 15 $\mu$eV energy resolution, we clearly discriminate PL lines corresponding to excitons in the electrostatic lattice with (0) to (4) inter-lattice NN interactions. These yield energy shifts of magnitude up to 4$\cdot V_{D,E}$. In the PL spectrum we model the emission from each of these states by assigning it with a profile given by our spectral resolution, at the corresponding energy. These contributions are referred to as lines (0)--(4) throughout the manuscript. In the dipolar potential intra-lattice NN couplings have an amplitude  reduced to $V_D\sim30-40$ $\mu$eV. In this situation, we possibly discriminate the contributions from each WS, WS1 and WS2 separated by 200 $\mu$eV. Also, in Fig.2.c we underline a contribution at around $2V_D$ from the one of WS1, corresponding to two intra-lattice NN couplings between excitons occupying WS1. On the other hand, we note that indications for NN couplings between excitons occupying WS2 are not clearly resolved. This possibly relates to the weak occupation of this high energy level at our bath temperature.}

\textcolor{black}{
In Fig.1.d, we quantitatively reproduce the PL profile by solely adjusting the weight of the lines (0)--(3) in the electrostatic lattice (red peaks), together with the contribution without intra-lattice NN couplings for WS1 (I) and WS2 (III) in the dipolar lattice (blue peaks). Each weight sets the average occupation of the corresponding state. To calibrate the filling $\bar{\nu}$ we measure the integrated intensity of the PL from our device. For each experiment we determine $\bar{\nu}$ by preparing a MI to ensure $\bar{\nu}=1$ and then we deduce the variation of the average filling by comparing the integrated intensity of the PL to this reference. For instance we thus conclude that $\bar\nu=1.2$ in Fig.1.c. }

\textcolor{black}{
In the regime where only few excitons occupy the dipolar lattice, i.e. for $\bar{\nu}\gtrsim1$, we can no longer rely on the computation of the integrated PL intensity to deduce the filling of the dipolar lattice, since this approach is not sufficiently precise. Instead we model the PL spectral profile from the electrostatic lattice. Figure S1 shows examples taken from the experiments reported in Fig.3.b. These spectra quantify the variation of the PL spectrum while the average filling is increased by weakly enhancing the laser excitation power,  starting from $\bar{\nu}\sim1$ (panels a) to c)). By modelling the PL spectra (red lines) we extract the amplitudes of all contributions, i.e. lines (0)-(4), which directly provide the fraction of excitons with at least one inter-lattice NN interaction.}

\newpage
 \vspace{.5cm}
\centerline{\includegraphics[width=\linewidth]{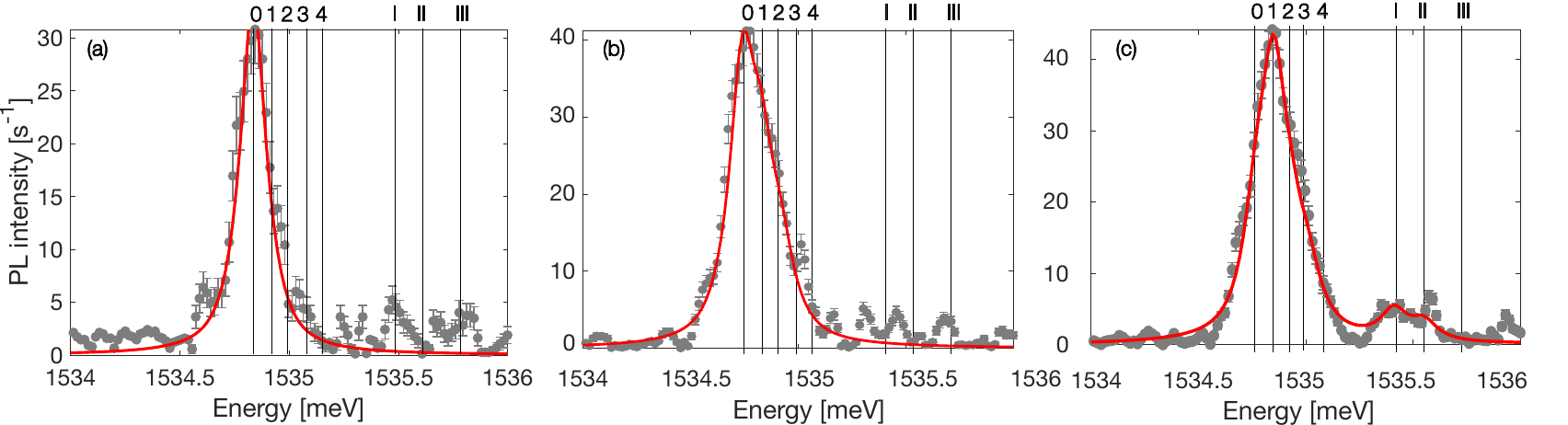}}
 \vspace{.5cm}\textcolor{black}{\textbf{Fig. S1: }
PL spectra slightly above unitary filling of the electrostatic lattice. The maximum of the spectrum shifts from the energy of line (0) in a), i.e. free from inter-lattice NN interactions, to the position of line (1) in the panel c), as $\bar{\nu}$ is slowly increased. Modelling the profiles (red lines) we deduce the fraction of excitons that experience at least 1 inter-lattice NN interaction. It is around 10$\%$ in a), 40$\%$ in b) and 75$\%$ in c). The normalised compressibility is equal to 0.45, 1 and 0.85 respectively. Displayed measurements are taken from the experiments reported in Fig.3.b.}
 \vspace{.5cm}

\section*{Stroboscopic measurements and exciton compressibility}

We perform photoluminescence (PL) stroboscopy. Precisely, we inject carriers in the electrostatic lattice by using a 100 ns long laser pulse repeated at a frequency of 1 MHz. The laser spot is (8x8) $\mathrm{\mu m}^2$ (1000 sites) wide and \textcolor{black}{obtained by stretching (defocusing) a gaussian excitation.}

The spatially and spectraly resolved PL is measured  in a 100 ns long time interval starting 300 ns after extinction of the laser pulse. \textcolor{black}{This ensures that dipolar excitons, with a lifetime around 700 ns, are studied at quasi-equilibrium and in the electro-neutral regime.
Indeed we do not observe any contribution due to charge excitons, unlike Ref.\,\cite{lagoin2023dual}, which ensures that the residual charges (holes) density in our device is bound to around \SI{4e7}{\per\centi \meter\squared}\cite{lagoin2023dual} compared to our excitons density around to \SI{2e9}{\per\centi \meter\squared} in the present experiments.}

 Let us note that our studies are characterised by weak PL emissions, so that integrations around 1 min long are necessary to measure a single spectrum with sufficient signal-to-noise-ratio. PL spectra are actually studied by recording repetitions for every experimental condition, subsequently averaged. Throughout the manuscript each spectrum represents an average over a few $10^8$ realisations. Using repetitions recorded in the same conditions, we further compute PL intensity statistical fluctuations to deduce the exciton compressibility $\kappa$. Indeed, the ratio between the standard deviation of the PL intensity and its mean expectation, $\sigma(I)/\bar{I}$, is directly proportional to $\sqrt{\kappa k_BT}$\, \cite{Gemelke2009}. The exciton compressibility normalised to its poissonian expectation in the electrostatic  lattice  $\kappa_E$, and in the dipolar lattice $\kappa_{D}$, are then inferred by monitoring intensity fluctuations of the corresponding local maxima of PL spectra.

\section*{Hubbard parameters in the electrostatic and dipolar- lattices}

We have previously shown in Ref.\,\cite{lagoin2022extended} that for the electrostatic lattice NN dipolar interactions have a strength $V_{E}= (30\pm5$) $\mu$eV, while the on-site interaction strength $U$ is theoretically around 1 meV. In the \SI{250}{\micro\eV} deep electrostatic lattice, excitons emulate then the Bose-Hubbard model extended by NN interactions and in the dipolar occupation blockade regime.

We deduce $V_{E,D}$ from the magnitude of $V_E$, since NN dipolar repulsions follow a $1/r^3$ scaling. Given the geometry of the two lattices we obtain $V_{E,D}=2\sqrt{2}\cdot V_{L}=$ 85 $\mu$eV. Let us note that we have implicitly assumed that excitonic wave-functions weakly vary between the electrostatic and dipolar- lattices. In this case, we calculate Hubbard parameters for the dipolar-lattice by considering its confinement depth  and the resulting profile of Wannier functions. Precisely, taking into account the depth of the electrostatic lattice $V_0$ and the strength of the dipolar potential, we estimate that the excitons potential energy at the position of the electrostatic barrier is around $(V_0+4V_{E,D})\sim 600$ $\mu$eV. To obtain two Wannier states in the dipolar-lattice potential separated by about 200 $\mu$eV, theoretically the dipolar-lattice depth is around 500 $\mu$eV. Hence, we recover the measured dipolar-lattice depth and that U is of the order of 1 meV. In this case, the tunnel parameters are 0.1 $\mu$eV (corresponding to a tunnel timescale around 6 ns) and 2.1$\mu$eV (0.3 ns) for the lower and upper energy WS respectively.

To confirm the previous estimations, we studied the energy of the PL as a function of the total filling factor $\bar{\nu}$. Figure S2.a first displays the energy of the contribution from the dipolar-lattice. We note that at low fillings the PL energy is the highest signalling that excitons mostly occupy the higher energy WS. The relaxation in the dipolar-lattice is then only partial which rules out the buildup of a quantum insulating phase. Let us then note that the energy relaxation in the dipolar-lattice is intrinsically correlated with the one in the electrostatic potential, since initially we optically inject high energy carriers in the gate defined electrostatic potential.

At  $\bar{\nu}$ around 3/2, we recover in Fig.S2.a that the PL energy suddenly decreases by around 200 $\mu$eV, as expected since excitons all occupy the lowest WS without intra-lattice NN interactions for CB order. Then, increasing $\bar{\nu}$ to around 1.8 leads to an increase of the PL energy by around 90 $\mu$eV. This enhancement corresponds to an average around 2 to 3 intra-lattice NN interactions, which confirms that $V_{D}$ is about 30 $\mu$eV. Moreover, Fig.S2.b displays the corresponding energy variation for excitons confined in the electrostatic lattice. Between lowest and highest fillings we note that the PL energy increases by around 200 $\mu$eV. This magnitude corresponds in average to an increase from 1NN to 3NN with excitons in the dipolar-lattice, confirming that $V_{E,D}$ is of the order of 90 $\mu$eV.\

The above magnitudes for $V_{D}$ and  $V_{E,D}$ were used to model PL spectra throughout the manuscript, along with a 200 $\mu$eV energy separation between the two WS of the dipolar lattice. On the other hand, we systematically imposed that the PL radiated by each state accessible in the electrostatic and dipolar lattices has a spectral profile  given by our spectrometer response function which is bound to 150 $\mu$eV. To model PL spectra we then only varied the weights of inter- and intra- lattices NN contributions (see for instance the insets in Fig. 2.b-c).

\centerline{\includegraphics[width=.8\linewidth]{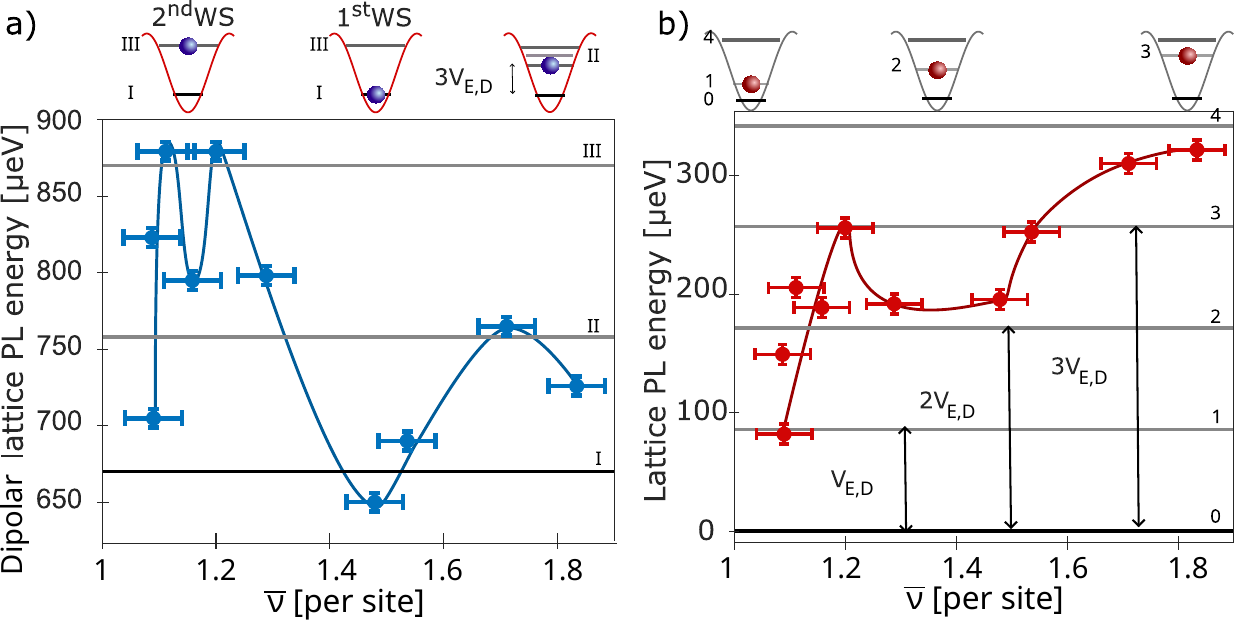}}\vspace{.2cm}
\textbf{Fig. S2: }
a) PL energy emitted by excitons in the dipolar-lattice as a function of the total filling factor  $\bar{\nu}$. Horizontal lines mark the energy of the second WS (top), the one of the lowest WS dressed by around 2 to 3 intra-lattice NN interactions (middle), and the bare lowest WS energy  (bottom). Top sketches illustrate the corresponding exciton confined levels. b) PL energy due to excitons confined in the electrostatic lattice as a function of  $\bar{\nu}$. Top sketches illustrate the corresponding exciton confined states as a function of the number of inter-lattice NN interactions. Horizontal lines display the PL energy shifts, from  0 to $4V_{E,D}$, starting from the bare lowest WS energy (black). Measurements have all been performed at $T=330$ mK and averaged around 50 lattice sites at the center of the optical injection.  

\end{document}